\documentclass[10pt,letterpaper]{article}

\usepackage[
  letterpaper,
  margin=1in,
  headheight=14pt,
  headsep=24pt,
  footskip=36pt
]{geometry}
\usepackage{amsmath,amssymb}
\usepackage{booktabs}
\usepackage{threeparttable}
\usepackage[T1]{fontenc}
\usepackage[utf8]{inputenc}
\usepackage{tgschola}
\usepackage{graphicx}
\usepackage{caption}
\usepackage{titlesec}
\usepackage{fancyhdr}
\usepackage[hidelinks,hyperfootnotes=false]{hyperref}
\usepackage[authoryear,round]{natbib}

\selectfont
\titleformat{\section}
  {\fontsize{13pt}{15.6pt}\selectfont\bfseries}
  {}{0pt}{}
\titleformat{\subsection}
  {\fontsize{11pt}{13.2pt}\selectfont\bfseries\itshape}
  {}{0pt}{}
\titleformat{\subsubsection}
  {\fontsize{10pt}{12pt}\selectfont\bfseries}
  {}{0pt}{}
\titlespacing*{\section}{0pt}{10pt}{10pt}
\titlespacing*{\subsection}{0pt}{10pt}{10pt}
\titlespacing*{\subsubsection}{0pt}{10pt}{10pt}

\title{Understanding Venture Capital Syndication in Information Technology Sectors: A Network Formation Perspective}
\newcommand{\paperauthors}{Liheng Tan, Zhengkai Tu\textsuperscript{*}, and Prasanna Karhade}
\newcommand{\paperaffiliation}{Department of Decisions, Operations and Technology, CUHK Business School}
\newcommand{\paperuniversity}{The Chinese University of Hong Kong, Hong Kong SAR}

\makeatletter
\renewcommand{\maketitle}{%
  \begin{center}
    {\fontsize{20pt}{24pt}\selectfont\bfseries \@title\par}
    \vspace{6pt}
    {\fontsize{12pt}{14.4pt}\selectfont\bfseries \paperauthors\par}
    \vspace{4pt}
    {\fontsize{10pt}{12pt}\selectfont \paperaffiliation\par}
    {\fontsize{10pt}{12pt}\selectfont \paperuniversity\par}
  \end{center}
  \begingroup
  \renewcommand{\thefootnote}{*}%
  \footnotetext{Corresponding author. Email: zhengkai.tu@cuhk.edu.hk}%
  \endgroup
}
\makeatother

\renewenvironment{abstract}
  {\section*{Abstract}\normalfont\normalsize}
  {\par}

\hypersetup{
  pdftitle={Understanding Venture Capital Syndication in Information Technology Sectors: A Network Formation Perspective},
  pdfauthor={Liheng Tan, Zhengkai Tu, Prasanna Karhade},
  pdfsubject={arXiv preprint},
  pdfkeywords={venture capital syndication, information technology sector, link formation, information asymmetry, exponential random graph models}
}

\begin{document}
\maketitle

\begin{abstract}
\noindent Venture capital syndication enables investors to pool diligence, share risk, and signal venture quality, while shaping the relationships through which investment networks develop. We examine how prior relationships, network embeddedness, and organizational similarity structure annual co-investment link formation in U.S. information technology venture finance. Using dyad-complete PitchBook panels for the hardware, software, and hybrid subsectors from 1966 to 2024, we test seven mechanisms through full-sample dyadic logit models with dyad-clustered standard errors. Across subsectors, prior collaboration is the most consistent correlate of co-investment; shared partners, geographic proximity, and organizational-type similarity are also positively associated with link formation, while domain overlap, prominence, and experience vary across settings. A static ERGM of the 2024 software network among 1,100 persistently active investors likewise produces positive estimates for triadic closure and geographic homophily and a smaller positive estimate for type homophily. By combining complete dyadic risk sets with a whole-network specification, the study shows how relational persistence, network closure, and homophily jointly structure IT venture syndication networks. In future work, we will extend the analysis with temporal network models, counterfactual simulations of market shocks, and evaluations of network-aware partner recommendations.
\end{abstract}


\section{Introduction}
\label{sec:intro}

Young IT firms often lack operating histories, tangible assets, and stable technical benchmarks, while much of their prospective value resides in code, data, architectures, or specialized human capital that outsiders cannot readily verify \citep{akerlof1978market,amit1998venture,aboody2000information}. Many develop through stages whose value is hard to benchmark in real time (e.g., blockchains; \citealp{li2026staging}). Venture investors must therefore evaluate both an uncertain startup and the judgment and follow-through of prospective co-investors. Syndication helps pool assessments and risk and draws information from credible participation \citep{bygrave1987syndicated,lockett2001syndication,stuart1999interorganizational}. Each syndicated deal also creates or renews a relationship, so repeated financing choices accumulate into a network that channels information and capital.

Research associates central VC-network positions with fund and portfolio outcomes \citep{hochberg2007whom,bellavitis2017impact,wu2024social} and uses relational and investor features to predict co-investment \citep{wang2015prediction,maus2024investor}. Formation studies identify reciprocity, community sorting, homophily, and closure as plausible mechanisms \citep{hochberg2010networking,bubna2020venture,gaonkar2023model,liu2021modeling,gao2025cross}. Yet these mechanisms have not been jointly assessed across major IT subsectors, and pairwise evidence may change when the outcome is represented as an interdependent network. A formation perspective therefore asks not simply whether centrality matters, but how repeated partner choices create the positions that later shape information and access. We treat network position as an accumulated outcome of partner selection rather than as an exogenous starting condition.

This focus motivates two questions. \textbf{RQ1 (Link-Formation Mechanisms):} Which relational, homophily, and prominence mechanisms are associated with annual co-investment links among active investor pairs, and how do they differ across hardware, software, and hybrid IT? \textbf{RQ2 (Network Interdependence):} For triadic closure, geographic homophily, and organizational-type homophily (H2--H4), how does inference from a software-sector ERGM compare with the dyadic-logit baseline?

We construct dyad-complete annual panels from PitchBook records of U.S. IT ventures, 1966--2024, and estimate subsector-specific logits on all eligible observations with dyad-clustered standard errors. The design makes the absence of a link an observed outcome for every pair that was simultaneously active, rather than a sampled comparison. We then estimate a static ERGM for the 2024 software network among 1,100 investors active throughout 2018--2024. Prior relationships are the most consistent correlate of renewed co-investment; shared partners and geographic and organizational similarity are also positive across subsectors, while domain overlap, prominence, and experience vary. In the ERGM, triadic closure and geographic homophily are positive and precisely estimated, whereas the positive type-homophily estimate is smaller and marginally significant.

The study contributes a unified formation framework for persistence, homophily, overlap, prominence, and experience in IT venture finance; distinguishes dyad-level associations from whole-network patterns; and informs digital partner-search systems that use relational features without equating predicted link likelihood with partner quality. The design supports descriptive and predictive inference, not causal claims, while establishing a basis for matched temporal and counterfactual network analysis.

\section{Theoretical Background and Hypotheses Development}
\label{sec:theory}

\subsection{VC Investment in IT: Sharper Information Asymmetry}

Venture investing is a canonical setting for decision-making under severe information asymmetry \citep{akerlof1978market}. Financiers must screen and monitor firms whose value depends on unobservable managerial quality, unproven technology, and intangible assets that are difficult to verify before investment \citep{sahlman2022structure,amit1998venture}. Unlike public-market investors, venture investors cannot rely on long operating histories, audited disclosures, analyst coverage, or continuous price discovery. Young firms offer few external benchmarks with which to evaluate either their prospects or the judgments of those proposing to finance them.

The uncertainty operates at two levels. Investors must assess the venture, but they must also assess one another. A prospective co-investor's technical judgment, diligence effort, access to capital, and willingness to provide follow-on support are themselves difficult to observe before collaboration \citep{gulati1995familiarity,gulati1999where}. Partner selection therefore creates an information problem layered on top of deal selection. Prior relationships, shared contacts, and third-party endorsements become valuable because they reveal qualities of a potential collaborator that formal deal materials cannot establish. Network links are consequently not only channels through which capital moves; they are also records of whom investors have previously trusted and indirect signals about whom others consider credible.

IT ventures sharpen both levels of uncertainty. Their value often resides in code, data, architectures, or specialized technical talent rather than readily auditable physical assets \citep{aboody2000information}. Technical quality may be difficult for outsiders to evaluate, and relevant benchmarks can change quickly as technologies and markets evolve. At the same time, IT venture activity is concentrated in a relatively small number of investment hubs \citep{sorenson2001syndication,chen2010buy,cumming2010local}. This concentration also operates in two directions: geographic proximity facilitates informal information exchange and repeated interaction, but it places credible collaborators and direct competitors in the same local market. Technical opacity and spatial concentration therefore make information about both ventures and prospective partners especially valuable and locally contested. In this setting, syndication is not merely a way to share a financing round; it is a mechanism through which investors select, evaluate, and repeatedly access collaborators.

\subsection{VC Syndication and Research Gap}

VC syndication is a long-standing organizational response to these information problems. It first enables information pooling: a lead investor can share technical assessments and due-diligence findings so that a group evaluates a venture more thoroughly than any participant could alone \citep{lockett2001syndication}. Second, it distributes capital commitments and downside exposure across multiple investors, easing both financing constraints and portfolio concentration risk \citep{bygrave1987syndicated}. Third, it creates reputational signals. When venture quality cannot be verified directly, participation by a credible investor can certify the opportunity to others who have less information \citep{stuart1999interorganizational}. A fourth mechanism addresses the partner side of the problem directly: working alongside a partner on a live deal reveals that partner's diligence and follow-through in a way reputation alone cannot, and this experience accumulates only through repeated collaboration with the same partner \citep{sorenson2001syndication}. These mechanisms are mutually reinforcing, but none operates independently of partner choice: the value of shared information, risk capacity, certification, and partner learning all depend on who joins the syndicate. Understanding syndication therefore requires explaining not only why investors collaborate, but also why a particular pair forms or renews a link.

Existing empirical research addresses this question from three related directions. A prediction-oriented strand uses investor and deal attributes to forecast which investor pairs will co-invest. Its predictors include reputation, prior relationships, and geographic proximity \citep{wang2015prediction,maus2024investor}. An outcomes-oriented strand examines what follows from network participation, linking an investor's position in the syndication network to fund performance and portfolio-company outcomes \citep{hochberg2007whom,bellavitis2017impact,wu2024social}. A formation-oriented strand investigates how the network itself emerges. This work shows, for example, that established VCs may use dense and reciprocal relationships to limit newcomers' access to deals \citep{hochberg2010networking}, that investors sort into persistent communities of similar partners \citep{bubna2020venture}, and that homophily and structural closure jointly generate observed network patterns \citep{gaonkar2023model,liu2021modeling}. Related network evidence also emphasizes the joint relevance of these mechanisms \citep{gao2025cross}.

Two gaps remain across these strands. First, studies of network consequences often begin with an investor's observed position as an explanatory condition. Yet position is not assigned independently of investor behavior; it accumulates through earlier partner choices made by firms with particular reputations, locations, specializations, and resources \citep{hochberg2007whom,bellavitis2017impact,wu2024social}. Without examining the formation process, it is difficult to distinguish the benefits associated with being well connected from the attributes and choices that make an investor well connected. A formation perspective therefore treats network position as an outcome to be explained before it is used to explain later performance.

Second, formation research often emphasizes a particular mechanism---such as exclusivity and reciprocity \citep{hochberg2010networking}, similarity-based community sorting \citep{bubna2020venture}, or structural closure \citep{gaonkar2023model,liu2021modeling,gao2025cross}---rather than comparing relational history, homophily, specialization, prominence, and experience within a common empirical design. Integrative evidence is especially limited for IT venture finance, where technical opacity and geographic concentration should make both prior relationships and network-based signals consequential. We address this gap by evaluating seven complementary mechanisms side by side across hardware, software, and hybrid IT subsectors. We then compare the dyadic results with a whole-network specification for the software network, allowing the evidence on closure and homophily to be assessed under two representations of network formation.

\subsection{Hypotheses}

The preceding arguments imply three complementary routes through which investors can manage uncertainty. Relational mechanisms use direct or indirect experience: prior links and shared partners lower the cost of assessing a prospective collaborator (H1--H2). Homophily mechanisms make a partner more legible through geographic, organizational, or domain similarity (H3--H5). Prominence and experience provide observable signals of standing, but may also increase selectivity or competition (H6--H7). We examine these mechanisms jointly because their associations with link formation need not be separable in a networked market.

\textbf{H1 (Relationship Persistence):} Prior co-investment links are positively associated with link formation: previously linked investor pairs are more likely to form a link in a given year than previously unlinked pairs \citep{maus2024investor}. Once two investors have co-invested, much of the information problem described at the start of this section is already resolved: each has direct experience of the other's diligence, judgment, and follow-through, which is exactly the kind of information that is otherwise so costly to obtain. Renewing a link economizes on this accumulated experience rather than paying the search and verification costs of a new partner.

\textbf{H2 (Triadic Closure / Structural Embeddedness):} Shared co-investment partners are positively associated with link formation: the probability of link formation increases with the number of co-investment partners two investors share \citep{wang2015prediction,gaonkar2023model}. Shared partners are a substitute for direct experience: even without a prior link of their own, two investors who have each already vetted the same third party can draw on that party's implicit endorsement, and are more likely to encounter each other in the same deals to begin with.

\textbf{H3 (Geographic Homophily):} Geographic co-location is positively associated with link formation: co-located investor pairs are more likely to form a link than pairs located in different states \citep{sorenson2001syndication,chen2010buy}. Co-location lowers the cost of the informal, ongoing monitoring that is otherwise scarce in venture investing: nearby investors can meet, observe, and verify each other more cheaply than distant ones.

\textbf{H4 (Type Homophily):} Organizational-type similarity is positively associated with link formation: investor pairs of the same organizational type are more likely to form a link than pairs of different types \citep{wang2015prediction,bubna2020venture}. Investors of the same organizational type tend to share underwriting standards, decision processes, and time horizons, which reduces the coordination costs of syndicating together and makes each party easier for the other to evaluate.

\textbf{H5 (Domain/Expertise Overlap):} Field overlap is hypothesized to have an unsigned association with link formation; the closest available VC-network estimate favors a positive association \citep{gao2025cross}. Overlapping industry focus could cut either way, signaling complementary expertise or competition for the same deals \citep{sorenson2001syndication}.

\textbf{H6 (Network Centrality / Status):} Network centrality is positively associated with partner attractiveness: more central investors are more likely to be chosen as co-investment partners \citep{hochberg2007whom}. A partner's position in the network is itself informative when the partner's own quality is hard to verify directly: a highly central investor has, by revealed preference, already been chosen by many others, a credible signal of quality that a newcomer cannot easily fake.

\textbf{H7 (Investor Experience):} Investor experience is hypothesized to have a non-monotonic association with partner attractiveness \citep{cumming2010local}. A longer track record is the most direct available signal of investor quality, but it also expands an investor's own opportunity set: a highly experienced investor has more potential partners to choose among, which can dilute the odds that any one specific pair links in a given year even as that investor's total number of links grows.

\textbf{RQ2 (Network Interdependence):} For H2--H4, how does inference change when network interdependence is represented directly? ERGMs jointly represent homophily and triadic closure rather than treating dyads as independent \citep{arora2025microfoundations,gaonkar2023model,liu2021modeling,gao2025cross}. We compare the software dyadic estimates with a static ERGM to examine H2--H4 under pairwise and whole-network representations.

\section{Data and Measurement}
\label{sec:data}

\subsection{Data and Network Construction}

\textbf{Data.} PitchBook records cover three IT groupings: \emph{software} (Software, IT Services), \emph{hardware} (Computer Hardware, Semiconductors), and \emph{hybrid} (Communications and Networking, Other Information Technology). We retain deals involving US-headquartered portfolio companies through 2024; investor nationality is unrestricted. Table~\ref{tab:subsector-scale} shows the much larger scale of software and that 68--74\% of deals in every subsector are syndicated.

\begin{table}[htbp]
\centering
\begin{tabular}{lrrr}
\toprule
 & Hybrid & Hardware & Software \\
\midrule
Deals & 3,384 & 5,812 & 67,034 \\
Date range & 1969--2024 & 1966--2024 & 1969--2024 \\
Distinct investors & 4,134 & 8,343 & 55,044 \\
Max active in one year & 742 & 1,399 & 14,216 \\
Syndicated deals & 73.8\% & 67.9\% & 70.4\% \\
\bottomrule
\end{tabular}
\caption{Deal and investor counts by IT subsector.}
\label{tab:subsector-scale}
\end{table}

\textbf{Procedure.} We construct two related objects for each subsector. First, a cumulative co-investment graph, in which an edge between two investors, once formed, persists in all subsequent years; this graph is the basis for each investor's network features (Section~\ref{sec:method}) as of year $t-1$. The cumulative representation retains the relationship history available to investors at the time of a new financing decision, while the outcome remains an annual co-investment link. Second, the set of investors \emph{active} in year $t$ (those making at least one deal that year) defines the analysis population: only active investors, and only pairs drawn from them, enter the panel for that year. Every such pair, in every year, generates one row of the panel (no negative sampling), with $\text{Link}_{ijt}=1$ if the pair co-invests that year.

\begin{figure}[htbp]
\centering
\includegraphics[width=\textwidth]{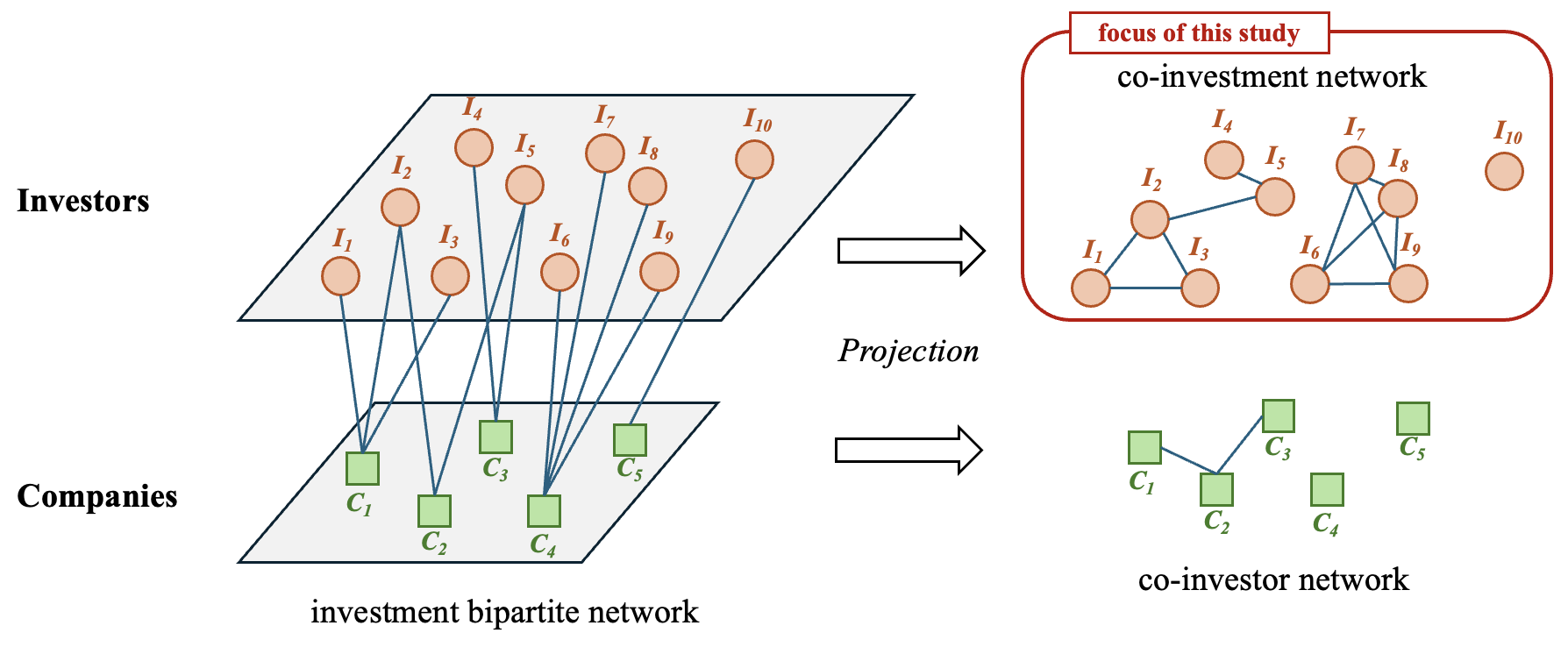}
\caption{Network projection used in this study. The boxed investor projection is the empirical object of interest: an annual co-investment edge links two active investors when they finance at least one common deal.}
\label{fig:network-construction}
\end{figure}

Figure~\ref{fig:network-construction} depicts the projection used in the analysis. The source records form a two-mode incidence structure linking investors to portfolio-company deals \citep[e.g.,][]{zhou2020phase}. We project that structure onto the investor layer, which preserves the partner relations relevant to syndication while abstracting from the individual deal nodes: two active investors are adjacent in year $t$ when they co-fund at least one deal that year.

\subsection{Variable Measurement}

\textbf{Variables.} The dependent variable, $\text{Link}_{ijt}$, is a binary indicator equal to 1 if investors $i$ and $j$ co-invest in at least one deal in year $t$. Table~\ref{tab:variables} lists the independent variables, one or more per hypothesis, and the controls. \texttt{CommonNeighborRatio} (H2) is the share of $i$ and $j$'s co-investment partners that are shared by both. \texttt{SameState} (H3) and \texttt{SameInvestorType} (H4) are binary indicators of shared HQ state and organizational type; when either investor's state or type is unreported, the pair is coded 0 rather than dropped, so these two indicators understate similarity to the extent the unreported cases include true matches. \texttt{FieldsJaccard} (H5) is the Jaccard similarity of $i$ and $j$'s historical industry-field portfolios. \texttt{Degree} and \texttt{Betweenness} (H6), and \texttt{LogDeals} (H7, log of one plus cumulative deal count), each enter as a Max/Min pair, the larger and smaller of the two investors' values, so the specification does not have to assume symmetry between the more and less prominent or experienced member of a pair.

\textbf{Timing.} Independent variables and controls are computed on the cumulative graph through $t-1$, preventing the current link from mechanically entering its own predictors but not establishing causal identification. \texttt{ShortestDistanceBin} records prior shortest-path distance (1, already linked, through 5+, or disconnected); other controls are a centered year trend and, in the pooled specification, subsector fixed effects.

\begin{table}[htbp]
\centering
\resizebox{\textwidth}{!}{%
\begin{tabular}{llll}
\toprule
Variable & Definition & Hypothesis & Hypothesis Sign \\
\midrule
CommonNeighborRatio & Share of shared co-investment partners & H2 & $+$ \\
SameState & =1 if same HQ state & H3 & $+$ \\
SameInvestorType & =1 if same organizational type & H4 & $+$ \\
FieldsJaccard & Jaccard similarity of industry-field portfolios & H5 & $\pm$ \\
Degree\_Max/Min & Larger/smaller network degree & H6 & $+$ \\
Betweenness\_Max/Min & Larger/smaller betweenness centrality & H6 & $+$ \\
LogDeals\_Max/Min & Larger/smaller log(1+cumulative deals) & H7 & $\pm$ \\
\midrule
ShortestDistanceBin & Shortest-path distance category (Distance=1 baseline) & H1 (control) & $-$ \\
Year (centered) & Linear year trend & --- & --- \\
Category FE & Subsector fixed effects (pooled spec.) & --- & --- \\
\bottomrule
\end{tabular}
}
\caption{Independent variables and controls, by hypothesis.}
\label{tab:variables}
\end{table}

\subsection{Descriptive Statistics}

Table~\ref{tab:descriptives} reports subsector means and standard deviations. Software has the largest and sparsest candidate-pair panel: its realized-link and common-neighbor shares are lowest, and missing state information is most prevalent. Previously linked pairs remain uncommon in every subsector (1.5\% in hybrid, 0.5\% in hardware, and 0.3\% in software).

\begin{table}[htbp]
\centering
\begin{tabular}{lrrr}
\toprule
 & Hybrid & Hardware & Software \\
\midrule
LogDeals\_Max & 1.473 (1.188) & 1.212 (1.098) & 1.741 (1.374) \\
LogDeals\_Min & 0.404 (0.678) & 0.253 (0.526) & 0.422 (0.738) \\
FieldsJaccard & 0.333 (0.471) & 0.157 (0.327) & 0.280 (0.425) \\
Degree\_Max & 27.353 (42.673) & 22.300 (41.042) & 81.482 (167.855) \\
Degree\_Min & 4.009 (11.032) & 2.433 (7.730) & 8.836 (29.172) \\
CommonNeighborRatio & 0.0070 (0.0291) & 0.0021 (0.0172) & 0.0019 (0.0095) \\
Betweenness\_Max & 0.0033 (0.0089) & 0.0018 (0.0065) & 0.0003 (0.0012) \\
Betweenness\_Min & 0.0002 (0.0012) & 0.0001 (0.0007) & 0.0000 (0.0001) \\
\addlinespace
Link & 0.0204 (0.1413) & 0.0100 (0.0996) & 0.0022 (0.0469) \\
SameState & 0.1638 (0.3701) & 0.1730 (0.3782) & 0.1484 (0.3555) \\
SameInvestorType & 0.3293 (0.4700) & 0.3131 (0.4638) & 0.2845 (0.4512) \\
\midrule
$N$ (SameState) & 1,173,807 & 3,801,446 & 229,384,568 \\
$N$ (all other rows) & 1,621,614 & 6,297,537 & 504,948,481 \\
\bottomrule
\end{tabular}
\caption{Descriptive statistics, dyad-complete panel variables, by subsector (mean, SD in parentheses).}
\label{tab:descriptives}
\end{table}

\section{Dyadic Logit: A Reduced-Form Baseline}
\label{sec:method}

\subsection{Model Specification}

The baseline is a dyad-wise logit that treats link probabilities as conditionally independent and does not model unobserved dyad heterogeneity. It provides a transparent reduced-form benchmark for assessing whether observed relational and investor attributes are associated with an annual link. We cluster standard errors by dyad because the same pair can recur across years; clustering corrects inference for repeated observations but not the point estimates.

For each subsector, we estimate the following model over all contemporaneously active investor pairs:

\begin{equation}
\begin{split}
\text{logit}\left(\Pr(\text{Link}_{ijt}=1)\right) = {}& \beta_0 + \beta_1\,\text{CommonNeighborRatio}_{ij,t-1} + \beta_2\,\text{SameState}_{ij} \\
& + \beta_3\,\text{SameInvestorType}_{ij} + \beta_4\,\text{FieldsJaccard}_{ij,t-1} \\
& + \mathbf{X}_{ij,t-1}\boldsymbol{\gamma} + \sum_{d} \delta_d\,\text{Dist}_{ij,t-1,d} \\
& + \gamma_Y\,\text{Year}_t + \varepsilon_{ijt},
\end{split}
\label{eq:linklogit}
\end{equation}

where $\mathbf{X}_{ij,t-1}$ contains the Max/Min pairs for degree, betweenness, and log cumulative deal count, and $\text{Dist}_{ij,t-1,d}$ are indicators for the pair's shortest network-path distance, with $d=1$ (already linked) as the omitted baseline. Estimation is separate by subsector so that common coefficients are not imposed on markets with different scale and competitive structure. Dyad clustering leaves the point estimates $\hat\beta$ unchanged while allowing observations of the same pair to be correlated across years.

\subsection{Results}

Table~\ref{tab:results-clustered} reports Equation~\ref{eq:linklogit} with dyad-clustered standard errors. Relative to nonrobust standard errors, clustering changes one substantive inference: hybrid's \texttt{Betweenness\_Min} is no longer significant ($p=0.117$); the remaining conclusions are unchanged.

\begin{table}[htbp]
\centering
\begin{threeparttable}
\normalsize
\begin{tabular}{lrrr}
\toprule
 & Hybrid & Hardware & Software \\
\midrule
CommonNeighborRatio & 3.272$^{***}$ (0.162) & 2.344$^{***}$ (0.113) & 5.458$^{***}$ (0.032) \\
SameState & 0.544$^{***}$ (0.018) & 0.557$^{***}$ (0.013) & 0.566$^{***}$ (0.003) \\
SameInvestorType & 0.233$^{***}$ (0.014) & 0.331$^{***}$ (0.010) & 0.605$^{***}$ (0.002) \\
FieldsJaccard & $-$0.427$^{***}$ (0.050) & 0.180$^{***}$ (0.026) & 0.109$^{***}$ (0.005) \\
Degree\_Max & $-$0.0030$^{***}$ (0.0003) & 0.0041$^{***}$ (0.0002) & 0.0007$^{***}$ (0.0000) \\
Degree\_Min & $-$0.0090$^{***}$ (0.0013) & $-$0.0002 (0.0009) & 0.0009$^{***}$ (0.0000) \\
Betweenness\_Max & 21.749$^{***}$ (0.731) & 4.962$^{***}$ (0.767) & 17.101$^{***}$ (0.385) \\
Betweenness\_Min & 13.527 (8.636) & $-$11.917$^{**}$ (4.799) & 7.273$^{***}$ (2.757) \\
LogDeals\_Max & $-$0.082$^{***}$ (0.011) & $-$0.188$^{***}$ (0.007) & 0.144$^{***}$ (0.001) \\
LogDeals\_Min & $-$0.210$^{***}$ (0.030) & $-$0.475$^{***}$ (0.023) & $-$0.127$^{***}$ (0.002) \\
Distance = 2 (ref.\ 1: already linked) & $-$2.337$^{***}$ (0.038) & $-$2.382$^{***}$ (0.032) & $-$1.660$^{***}$ (0.006) \\
Distance = 3 & $-$3.072$^{***}$ (0.048) & $-$3.041$^{***}$ (0.037) & $-$2.685$^{***}$ (0.007) \\
Distance = 4 & $-$3.327$^{***}$ (0.064) & $-$3.518$^{***}$ (0.045) & $-$3.176$^{***}$ (0.011) \\
Distance = 5+ & $-$3.077$^{***}$ (0.099) & $-$3.711$^{***}$ (0.063) & $-$3.247$^{***}$ (0.028) \\
Disconnected & $-$3.540$^{***}$ (0.063) & $-$3.494$^{***}$ (0.043) & $-$2.701$^{***}$ (0.008) \\
Year (centered) & 0.043$^{***}$ (0.001) & $-$0.037$^{***}$ (0.001) & $-$0.061$^{***}$ (0.000) \\
Constant & $-$0.619$^{***}$ (0.065) & $-$1.350$^{***}$ (0.044) & $-$4.451$^{***}$ (0.009) \\
\midrule
$N$ & 1,297,291 & 5,038,029 & 504,948,481 \\
Pseudo $R^2$ & 0.096 & 0.072 & 0.106 \\
\bottomrule
\end{tabular}
\caption{Determinants of co-investment, dyad-complete panel, dyad-clustered standard errors.}
\label{tab:results-clustered}
\begin{tablenotes}
\normalsize
\item $^{*}p<0.10$, $^{**}p<0.05$, $^{***}p<0.01$.
\end{tablenotes}
\end{threeparttable}
\end{table}

The clearest result in Table~\ref{tab:results-clustered} is H1. Distance = 1, already-linked pairs, is the omitted baseline, so every distance coefficient is a comparison against it, and all are large and negative: a distance-2 pair has only 9--19\% of the odds of an already-linked pair ($\exp(-2.337)=0.097$ hybrid, $\exp(-2.382)=0.092$ hardware, $\exp(-1.660)=0.190$ software), and the gap widens further at distance 3 and beyond. Prior relationships are renewed at a far higher rate than new ones form among unlinked pairs, in all three subsectors.

H2, H3, and H4 hold with the predicted sign and are significant in every subsector: shared partners (CommonNeighborRatio), co-location (SameState), and shared organizational type (SameInvestorType) all raise the odds of co-investment, with software showing the largest coefficients on all three.

The remaining hypotheses show real heterogeneity across subsectors rather than a uniform pattern. FieldsJaccard (H5) is negative in hybrid but positive in hardware and software, consistent with H5's two-sided prediction: shared domain focus can mean complementary expertise or direct competition, and which one dominates appears to differ by subsector. Degree (H6) is negative on both ends in hybrid, essentially null on the low end in hardware, and positive on both ends in software. LogDeals (H7) splits within software itself: LogDeals\_Max is positive while LogDeals\_Min is negative, exactly the pattern H7 anticipates when a highly active investor's own opportunity set dilutes the odds of any one specific pair, whereas hybrid and hardware are negative on both ends. Model fit, measured by pseudo $R^2$, is highest in software (0.106) and lowest in hardware (0.072).

\section{ERGM: Accounting for Network Interdependence}
\label{sec:ergm}

\subsection{Model Specification}

The dyadic logit treats links as conditionally independent even though shared-neighbor and centrality measures derive from the whole network. An ERGM instead assigns probability to an entire network configuration and can represent nonlinear transitivity directly:
\begin{equation}
\Pr(G=g\mid\theta) = \frac{\exp\left(\theta^\top s(g)\right)}{Z(\theta)}, \qquad Z(\theta) = \sum_{g'\in\mathcal{G}} \exp\left(\theta^\top s(g')\right),
\label{eq:ergm}
\end{equation}
where $s(g)$ is a vector of network-level statistics, rather than factoring the likelihood as a product over independent dyads. In particular, the GWESP term captures the diminishing contribution of additional shared partners to closure, rather than treating shared-partner exposure solely as a linear dyadic covariate.

We estimate a static ERGM for the 2024 software network among 1,100 investors active in every year from 2018 through 2024 (8,649 edges; density 1.4\%). Restricting the network to persistently active investors focuses the analysis on durable participants. The specification includes edges, GWESP for transitivity (fixed decay 0.1), and \texttt{nodematch} terms for HQ state and investor type, representing H2--H4 but not H1 or H5--H7. Estimation uses MCMLE with seed 42, an MCMC sample size of 4,000, burn-in of 100,000, interval of 2,000, and a maximum of 20 MCMLE iterations.

\subsection{Results}

\begin{table}[htbp]
\centering
\begin{threeparttable}
\normalsize
\begin{tabular}{lr}
\toprule
 & Software (2024) \\
\midrule
Edges & $-$12.067$^{***}$ (0.239) \\
GWESP (decay = 0.1) & 7.498$^{***}$ (0.206) \\
Same HQ State & 0.351$^{***}$ (0.019) \\
Same Investor Type & 0.029$^{*}$ (0.016) \\
\midrule
$N$ (investors) & 1,100 \\
\bottomrule
\end{tabular}
\caption{Static ERGM, software co-investment network, 2024, persistently active investors.}
\label{tab:ergm-results}
\begin{tablenotes}
\normalsize
\item $^{*}p<0.10$, $^{**}p<0.05$, $^{***}p<0.01$.
\end{tablenotes}
\end{threeparttable}
\end{table}

Table~\ref{tab:ergm-results} answers RQ2 at the whole-network level. GWESP is positive and significant, indicating that links are concentrated in locally closed structures with shared partners. Same HQ state is also positive and significant, consistent with geographic homophily. Same investor type is positive but smaller and marginally significant ($p=0.065$), indicating that organizational similarity has a more modest association once transitivity and geographic matching are represented jointly.

\section{Discussion}
\label{sec:discussion}

With respect to RQ1, co-investment formation reflects several mechanisms rather than one universal rule. Relationship persistence (H1) is the most stable cross-subsector result: prior partners are consistently more likely to invest together again. This supports the view of an existing link as relational capital that reveals how a partner evaluates technology, coordinates, and responds under uncertainty, thereby lowering later search and verification costs \citep{lockett2001syndication,maus2024investor}. Structural embeddedness, geographic proximity, and organizational-type similarity (H2--H4) also point in the predicted directions in the dyadic analysis. These mechanisms are distinct: shared partners can provide referrals, proximity can ease communication and monitoring, and organizational similarity can make routines and horizons more predictable \citep{sorenson2001syndication,chen2010buy,gaonkar2023model}.

The mixed evidence for domain overlap, prominence, and experience (H5--H7) shows that partner attractiveness is contextual and relational. Similar expertise may facilitate joint evaluation but also intensify competition; prominence and experience may signal quality while increasing selectivity or expanding the opportunity set \citep{hochberg2007whom,cumming2010local}. Thus, the same investor characteristic can affect matching differently across subsectors and depending on which member of the dyad possesses it. Theoretically, this shifts attention from treating network position only as a cause of performance to examining how repeated partnering, embeddedness, and selective matching produce that position in the first place \citep{bellavitis2017impact,wu2024social}.

RQ2 receives a complementary whole-network answer. The software ERGM aligns with the dyadic logit on triadic closure and geographic homophily, while organizational-type homophily is smaller in the network specification. The positive GWESP term shows that shared-partner structures are strongly associated with co-investment, and the positive state term indicates that geography remains relevant when these structures are modeled jointly \citep{gaonkar2023model,liu2021modeling,gao2025cross}. The two representations therefore provide consistent evidence for embeddedness and proximity while showing that the magnitude of type homophily depends on model representation.

Practically, VC managers can use prior collaboration and shared contacts as screening signals while guarding against closed-circle reinforcement. Entrepreneurs should expect early investor choices to shape later syndication opportunities, but should not assume that the most central investor is always the best match. Co-investor-matching platforms can incorporate relationship history, shared contacts, geography, and organizational type while separating predicted link likelihood from partner quality and auditing whether recommendations amplify incumbent visibility.

Overall, IT co-investment positions accumulate through repeated links, embeddedness, similarity, and selection that varies by context. The dyadic models document these mechanisms across subsectors and years, while the ERGM complements them with a whole-network representation of closure and homophily. Together, the results identify which mechanisms are consistently associated with formation and which vary with sector and model representation.

The combined results also clarify an important interpretive boundary. Strong renewal and closure associations imply that access is path dependent: existing relationships create information and opportunities for later relationships. They do not establish that repeated collaboration improves deal quality or that a highly connected investor is the best partner for every venture. Network-aware decision systems should consequently treat relational features as signals of feasibility and information access, while evaluating whether their use changes the visibility of entrants or reinforces concentration among incumbent investors.

\section{Limitations and Future Research}
\label{sec:limitations}

Several limitations qualify these findings. The covariate set omits deal size and valuation, lead-investor status, contemporaneous syndicate size, and LP-level relationships; the cumulative network weights old and recent links equally; the industry-overlap measure relies on a coarse taxonomy; and missing location or investor-type data is coded as non-matching, all of which may obscure recency, domain specificity, and homophily. The design is descriptive and predictive rather than causal: dyad-clustered standard errors address repeated observations but not bias from unobserved dyad-specific factors, so H1--H7 should not be read as causal effects. The ERGM describes a single observed network with parameters that have not yet been systematically fine-tuned, so the analysis cannot trace formation and dissolution as evolving processes or speak to how the network responds to a shock. Finally, the single-country, single-sector setting is an intentional boundary condition that limits generalizability to other institutional and technological environments.

\subsection{Priority Directions for Future Research}

First, a temporal ERGM (TERGM) should model yearly network evolution using aligned node populations and formation and dissolution processes. Beyond dynamic estimation, the fitted model can support counterfactual simulations that introduce shocks -- such as investor exits, fund closures, relocations, or regional disruptions -- and compare their effects on link persistence, network concentration, and access to syndication partners against a no-shock baseline. A dynamic design can also test whether relationships weaken with age rather than receiving the equal cumulative weight used here.

Second, future work should strengthen measurement and identification by adding deal- and fund-level covariates, applying time-decay or rolling-window network measures, treating missing homophily data explicitly, and incorporating correlated random effects \citep{arora2025microfoundations}. Candidate pairs should also be defined using only information available at the time a partner could realistically be considered. Where credible, quasi-experimental shocks can complement TERGM simulations with stronger causal evidence.

Third, external and predictive validity should be tested across countries and technology sectors and through out-of-time link prediction. Evaluation should report ranking accuracy, calibration, and changes in exposure for new, peripheral, and geographically distant investors. Such work can determine whether network-aware recommendations improve partner discovery without reinforcing concentration or excluding new investors.

Together, these extensions would move the study from a cross-sectional description of accumulated links toward evidence on how network structure evolves, how it changes under shocks, and how digital partner-search tools can be designed responsibly.

\bibliographystyle{apalike}
\bibliography{ref}

@article{wang2015prediction,
  title={The prediction of venture capital co-investment based on structural balance theory},
  author={Wang, Zhiyuan and Zhou, Yun and Tang, Jie and Luo, Jar-Der},
  journal={IEEE Transactions on Knowledge and Data Engineering},
  volume={28},
  number={2},
  pages={537--550},
  year={2015},
  publisher={IEEE}
}

@article{arora2025microfoundations,
  title={Microfoundations of link formation In supply networks},
  author={Arora, Kashish and Osadchiy, Nikolay},
  year={2025}
}

@article{bubna2020venture,
  title={Venture capital communities},
  author={Bubna, Amit and Das, Sanjiv R and Prabhala, Nagpurnanand},
  journal={Journal of Financial and Quantitative Analysis},
  volume={55},
  number={2},
  pages={621--651},
  year={2020},
  publisher={Cambridge University Press}
}

@article{gaonkar2023model,
  title={A model of inter-organizational network formation},
  author={Gaonkar, Shweta and Mele, Angelo},
  journal={Journal of Economic Behavior \& Organization},
  volume={214},
  pages={82--104},
  year={2023},
  publisher={Elsevier}
}

@article{liu2021modeling,
  title={Modeling venture capital networks in hospitality and tourism entrepreneurial equity financing: An exponential random graph models approach},
  author={Liu, Bing and Luo, Chaoliang and Meng, Fang and Jiang, Hui},
  journal={International Journal of Hospitality Management},
  volume={95},
  pages={102936},
  year={2021},
  publisher={Elsevier}
}

@article{gao2025cross,
  title={Cross-Layer Influence of Multiple Network Embedding on Venture Capital Networks in China: An ERGM-Based Analysis},
  author={Gao, Yuge and Xie, Yongping and Yang, Yanping},
  journal={Systems},
  volume={13},
  number={11},
  pages={1035},
  year={2025},
  publisher={MDPI}
}

@article{chen2010buy,
  title={Buy local? The geography of venture capital},
  author={Chen, Henry and Gompers, Paul and Kovner, Anna and Lerner, Josh},
  journal={Journal of Urban Economics},
  volume={67},
  number={1},
  pages={90--102},
  year={2010},
  publisher={Elsevier}
}

@article{cumming2010local,
  title={Local bias in venture capital investments},
  author={Cumming, Douglas and Dai, Na},
  journal={Journal of empirical finance},
  volume={17},
  number={3},
  pages={362--380},
  year={2010},
  publisher={Elsevier}
}

@article{sorenson2001syndication,
  title={Syndication networks and the spatial distribution of venture capital investments},
  author={Sorenson, Olav and Stuart, Toby E},
  journal={American journal of sociology},
  volume={106},
  number={6},
  pages={1546--1588},
  year={2001},
  publisher={The University of Chicago Press}
}

@article{hochberg2010networking,
  title={Networking as a barrier to entry and the competitive supply of venture capital},
  author={Hochberg, Yael V and Ljungqvist, Alexander and Lu, Yang},
  journal={The Journal of Finance},
  volume={65},
  number={3},
  pages={829--859},
  year={2010},
  publisher={Wiley Online Library}
}

@article{hochberg2007whom,
  title={Whom you know matters: Venture capital networks and investment performance},
  author={Hochberg, Yael V and Ljungqvist, Alexander and Lu, Yang},
  journal={The journal of finance},
  volume={62},
  number={1},
  pages={251--301},
  year={2007},
  publisher={Wiley Online Library}
}

@article{bellavitis2017impact,
  title={The impact of investment networks on venture capital firm performance: A contingency framework},
  author={Bellavitis, Cristiano and Filatotchev, Igor and Souitaris, Vangelis},
  journal={British Journal of Management},
  volume={28},
  number={1},
  pages={102--119},
  year={2017},
  publisher={Wiley Online Library}
}

@article{wu2024social,
  title={The social structure of insiders and outsiders: Toward a network community perspective on firm performance},
  author={Wu, Xiaoteng and Adbi, Arzi and Mahmood, Ishtiaq Pasha},
  journal={Academy of management journal},
  volume={67},
  number={4},
  pages={903--932},
  year={2024},
  publisher={Academy of Management Valhalla, NY}
}

@incollection{akerlof1978market,
  title={The market for ``lemons'': Quality uncertainty and the market mechanism},
  author={Akerlof, George A},
  booktitle={Uncertainty in economics},
  pages={235--251},
  year={1978},
  publisher={Elsevier}
}

@incollection{sahlman2022structure,
  title={The structure and governance of venture-capital organizations},
  author={Sahlman, William A},
  booktitle={Venture capital},
  pages={3--51},
  year={2022},
  publisher={Routledge}
}

@article{amit1998venture,
  title={Why do venture capital firms exist? Theory and Canadian evidence},
  author={Amit, Raphael and Brander, James and Zott, Christoph},
  journal={Journal of business Venturing},
  volume={13},
  number={6},
  pages={441--466},
  year={1998},
  publisher={Elsevier}
}

@article{aboody2000information,
  title={Information asymmetry, R\&D, and insider gains},
  author={Aboody, David and Lev, Baruch},
  journal={The journal of Finance},
  volume={55},
  number={6},
  pages={2747--2766},
  year={2000},
  publisher={Wiley Online Library}
}

@article{lockett2001syndication,
  title={The syndication of venture capital investments},
  author={Lockett, Andy and Wright, Mike},
  journal={Omega},
  volume={29},
  number={5},
  pages={375--390},
  year={2001},
  publisher={Elsevier}
}

@article{bygrave1987syndicated,
  title={Syndicated investments by venture capital firms: A networking perspective},
  author={Bygrave, William D},
  journal={Journal of Business Venturing},
  volume={2},
  number={2},
  pages={139--154},
  year={1987},
  publisher={Elsevier}
}

@article{stuart1999interorganizational,
  title={Interorganizational endorsements and the performance of entrepreneurial ventures},
  author={Stuart, Toby E and Hoang, Ha and Hybels, Ralph C},
  journal={Administrative science quarterly},
  volume={44},
  number={2},
  pages={315--349},
  year={1999},
  publisher={SAGE Publications}
}

@article{maus2024investor,
  title={How do investor characteristics of business angels and venture capitalists predict the occurrence of co-investments?},
  author={Maus, Christoph and Greven, Andrea and Kurth, Niklas and Brettel, Malte},
  journal={Journal of Business Economics},
  volume={94},
  number={5},
  pages={763--811},
  year={2024},
  publisher={Springer}
}

@article{zhou2020phase,
  title={Phase transitions and optimal algorithms for semisupervised classifications on graphs: From belief propagation to graph convolution network},
  author={Zhou, Pengfei and Li, Tianyi and Zhang, Pan},
  journal={Physical Review Research},
  volume={2},
  number={3},
  pages={033325},
  year={2020},
  publisher={APS}
}

@article{gulati1995familiarity,
  title={Does familiarity breed trust? The implications of repeated ties for contractual choice in alliances},
  author={Gulati, Ranjay},
  journal={Academy of Management Journal},
  volume={38},
  number={1},
  pages={85--112},
  year={1995}
}

@article{gulati1999where,
  title={Where do interorganizational networks come from?},
  author={Gulati, Ranjay and Gargiulo, Martin},
  journal={American journal of sociology},
  volume={104},
  number={5},
  pages={1439--1493},
  year={1999}
}

@misc{li2026staging,
  title={Staging the Development of Blockchain Platforms},
  author={Li, Tianyi and Tan, Liheng},
  year={2026},
  note={Available at SSRN 6665800}
}

\end{document}